\documentclass{article}
\usepackage{amsmath,graphicx,mlspconf}
\usepackage{amssymb,epsfig, url}
\usepackage{color,soul}
\usepackage{tcolorbox}
\usepackage{bbm}
\usepackage{hyperref}
\usepackage{cite}
\usepackage[font=small]{caption}
\usepackage{enumitem}
\usepackage[hang,flushmargin]{footmisc} 
\usepackage{framed}

\newcommand {\myvec}[1] {{\mbox{\boldmath $#1$}}}
\newcommand {\mymat}[1]  {{\mbox{\boldmath $#1$}}}

\newcommand {\us} {\myvec{s}}
\newcommand {\ub} {\myvec{b}}
\newcommand {\uy} {\myvec{y}}

\newcommand {\uz} {\myvec{z}}
\newcommand {\Cset} {\mathbb{C}}
\newcommand {\Rset} {\mathbb{R}}

\newcommand {\her} {\rm{H}}

\newcommand{\optarg}[1][]{%
  \ifthenelse{\isempty{#1}}%
    {}% if #1 is empty
    {(((#1)))}% if #1 is not empty
}

\newcommand {\Expec}[1] {\mathbb{E}\left[#1\right]}
\newcommand {\Expecto}[2] {\mathbb{E}_{#2}\left[#1\right]}
\newcommand {\CNorm} {\mathcal{C}\mathcal{N}}

\newcommand {\ust} {\us_{\tau_s}}
\newcommand {\ustest} {\widehat{\us}_{\tau_s}}
\newcommand {\ustestsub}[1] {\widehat{\us}_{\tau_s,#1}}
\newcommand {\ubt} {\ub_{\tau_b}}
\newcommand {\MMSE} {\text{\tiny MMSE}}
\newcommand {\LMMSE} {\text{\tiny LMMSE}}

\makeatletter
\newsavebox\myboxA
\newsavebox\myboxB
\newlength\mylenA

\newcommand*\mybar[2][0.75]{%
	\sbox{\myboxA}{$\m@th#2$}%
	\setbox\myboxB\null% Phantom box
	\ht\myboxB=\ht\myboxA%
	\dp\myboxB=\dp\myboxA%
	\wd\myboxB=#1\wd\myboxA% Scale phantom
	\sbox\myboxB{$\m@th\overline{\copy\myboxB}$}%  Overlined phantom
	\setlength\mylenA{\the\wd\myboxA}%   calc width diff
	\addtolength\mylenA{-\the\wd\myboxB}%
	\ifdim\wd\myboxB<\wd\myboxA%
	\rlap{\hskip 0.5\mylenA\usebox\myboxB}{\usebox\myboxA}%
	\else
	\hskip -0.5\mylenA\rlap{\usebox\myboxA}{\hskip 0.5\mylenA\usebox\myboxB}%
	\fi}
\makeatother
\let\OLDthebibliography\thebibliography
\renewcommand\thebibliography[1]{
  \OLDthebibliography{#1}
  \setlength{\parskip}{0pt}
  \setlength{\itemsep}{0pt plus 0.3ex}
}

\graphicspath{{figures/}}

\copyrightnotice{
\begin{minipage}{0.995\textwidth}
\begin{framed}\vspace{-0.5em}
\footnotesize {\copyright}2022 IEEE. Personal use of this material is permitted.  Permission from IEEE must be obtained for all other uses, in any current or future media, including reprinting/republishing this material for advertising or promotional purposes, creating new collective works, for resale or redistribution to servers or lists, or reuse of any copyrighted component of this work in other works. %(To appear in IEEE MLSP 2022) %DOI: \href{<To be added>}{To be added}
\vspace{-0.8em}
\end{framed}
\end{minipage}
}
\title{Exploiting Temporal Structures of Cyclostationary Signals for Data-Driven Single-Channel Source Separation}
\name{\parbox{\textwidth}{\centering Gary C.F. Lee, Amir Weiss, Alejandro Lancho, Jennifer Tang, Yuheng Bu, \\ Yury Polyanskiy, Gregory W. Wornell}
}
\address{Massachusetts Institute of Technology, Cambridge, MA}

\begin{document}\sloppy
\maketitle
\begin{abstract}
We study the problem of single-channel source separation (SCSS), and focus on cyclostationary signals, which are particularly suitable in a variety of application domains. Unlike classical SCSS approaches, we consider a setting where only examples of the sources are available rather than their models, inspiring a data-driven approach. For source models with underlying cyclostationary Gaussian constituents, we establish a lower bound on the attainable mean squared error (MSE) for any separation method, model-based or data-driven. Our analysis further reveals the operation for optimal separation and the associated implementation challenges. As a computationally attractive alternative, we propose a deep learning approach using a U-Net architecture, which is competitive with the minimum MSE estimator. We demonstrate in simulation that, with suitable domain-informed architectural choices, our U-Net method can approach the optimal performance with substantially reduced computational burden.
\end{abstract}

{\let\thefootnote\relax\footnotetext{ 
Research was sponsored by the United States Air Force Research Laboratory and the United States Air Force Artificial Intelligence Accelerator and was accomplished under Cooperative Agreement Number FA8750-19-2-1000. The views and conclusions contained in this document are those of the authors and should not be interpreted as representing the official policies, either expressed or implied, of the United States Air Force or the U.S. Government. The U.S. Government is authorized to reproduce and distribute reprints for Government purposes notwithstanding any copyright notation herein.\\
The authors acknowledge the MIT SuperCloud and Lincoln Laboratory Supercomputing Center for providing HPC resources that have contributed to the research results reported within this paper.\\
Alejandro Lancho is funded by EU’s Horizon 2020 programme under the Marie Sklodowska-Curie grant agreement No.~101024432.
}}

\section{Introduction}\label{sec:intro}
\vspace{-0.2cm}
Source separation is a well-studied problem with many important applications in radio-frequency systems, wireless communications and biomedical signal monitoring \cite{akeret2017radio, schlezinger2014fresh, zhao2021scbss, chen2019eeg}, to name a few. A common formulation is the blind source separation problem, for which perhaps the most popular framework is independent component analysis \cite{comon2010handbook}. Many approaches rely on some degree of spatial diversity present in multi-channel observations. In contrast, a particularly challenging problem of interest is the \emph{single-channel} (or single-sensor) source separation (SCSS). The goal in SCSS is to separate the latent sources from a single-channel observed mixture. In this regime, the aforementioned algorithms are irrelevant due to the absence of spatial diversity. Instead, other characteristics of the latent sources must be exploited. 

Recent efforts demonstrate that machine learning (ML) techniques can serve as powerful tools for source separation, even in the single-channel regime, in image and audio counterparts \cite{stoller2018wave,nugraha2016multichannel,gandelsman2019double,lyu2019reflection}. Solutions typically exploit the inherent structure specific to the signal type. For example, natural images may be separable by color features and local dependencies \cite{gandelsman2019double}, whereas speech signals are commonly addressed by time-frequency spectrogram masking \cite{huang2015joint,luo2019conv,tzinis2020sudo}. Furthermore, for time series ($1$-dimensional) data, if the sources are separable in time and/or frequency, appropriate spectrogram-based masking and classical filtering methods can be adopted, e.g. \cite{amin1997interference,amin2017time}. In contrast, a key challenge is the separation of signals that overlap (partially/fully) in both time and frequency; such a setting is rarely addressed in aforementioned works. Further, in such a regime, it is no longer trivial to identify the properties that are helpful in separating the signals, and what separation performance is attainable.

One particular type of time series data is the class of \emph{cyclostationary} signals, a relevant model adopted in many of the earlier mentioned applications (e.g., \cite{punchihewa2010cyclostationarity}). Interestingly, under some conditions, perfect separation of cyclostationary signals could be achieved despite having components with overlapping time-frequency spectra \cite{giannakis1998cyclostationary,gardner1994cyclostationarity}. The challenge, however, is in modeling such cyclostationarities in a way that properly describes the true statistics of the observed mixtures. A more realistic scenario is one where the signal model is unknown, but examples of the signals (by measurements or generation\footnote{Note that the ability to generate a signal (with some device) does not necessarily mean that one can also analytically characterize its true statistics.}) are available. Further, despite cyclostationarity properties, the available examples are usually finite unsynchronized time segments, meaning that they are extracted at possibly random different ``start times'', relative to an arbitrarily chosen time instance. 
In other words, there are additional latent variables---the random time shifts---that are unobservable, adding a layer of difficulty to the problem.

In this paper, we study the SCSS problem involving mixtures of two signals. Specifically, we focus on the scenario where each component is a randomly time-shifted and scaled segment from a cyclostationary complex Gaussian (hereafter, simply referred to as \emph{Gaussian} in this work) process, a regime in which an analytical form of the optimal estimator, in the sense of minimum mean-square-error (MMSE), can be derived. Nonetheless, we also highlight the challenges in implementing the optimal estimator, thereby motivating less computationally demanding alternatives. We propose a deep learning (DL) approach to the SCSS problem, and demonstrate its performance in two representative examples---one of which is based on real applications in wireless communications. Our simulations show that the performance of our proposed DL strategy is competitive to the optimal estimator.

We recognize that many signals of practical relevance generally depart from the Gaussian model. However, the analysis involving cyclostationary Gaussian signals allows us to better understand and characterize the performance of DL-based methods. In particular, it focuses our attention to the exploitation of temporal correlation structures of the latent source components, so as to accurately assess the ability of ML approaches in capturing such structures. This would shed light as to how far we are from the optimal performance, and how to close that performance gap.

\vspace{-0.4cm}
\section{SCSS Problem Formulation}\label{sec:prob}
\vspace{-0.25cm}

We consider the following model of an observed 1D signal of length $N$, which is a noisy mixture of two latent sources, 
{\setlength{\belowdisplayskip}{5pt} \setlength{\belowdisplayshortskip}{5pt}
\setlength{\abovedisplayskip}{5pt} \setlength{\abovedisplayshortskip}{5pt}
\begin{equation}
    \uy = \ust + \underbrace{\kappa \, \ubt}_{\text{``interference''}} + \underbrace{\sigma\uz}_{\text{``noise''}} \; \in \Cset^{N}, \label{mixture}
\end{equation}
}where $\ust, \ubt$ are the (unobservable) independent components, $\kappa\in\Rset_+$ is distributed according to some unknown (and for simplicity, discrete) distribution on $\mathcal{K}\subset\Rset_+$, $\myvec{0}$ denotes the all zeros-vector and $\mymat{I}$ denotes the identity matrix (with context-dependent dimensions). The noise component, $\sigma\uz$, is an additive white Gaussian noise, where $\uz \sim \CNorm(\myvec{0}, \mymat{I})$ and $\sigma^2$ corresponds to variance of the noise.
Without loss of generality, for the purposes of our discussion, $\ust$ is termed the ``reference'' signal and $\kappa \, \ubt$ the interference. Additionally, we assume that $\ust$ and $\ubt$ have unit average power; hence, $\kappa^2$ is related to the inverse of the signal-to-interference ratio (SIR). The goal in SCSS is to produce an estimate $\ustest$ based on $\uy$ so that given some metric $d$, the value $\Expec{d(\ustest, \ust)}$ is minimized. We limit our scope to the conventional metric $d(\myvec{u},\myvec{v})=\|\myvec{u}-\myvec{v}\|^2_2$, leading to the MMSE criterion.

We focus our discussion on signal components that are segments extracted from cyclostationary Gaussian processes. We consider two independent, discrete-time, zero-mean circularly-symmetric complex Gaussian processes $\tilde{s}[\cdot]$, $\tilde{b}[\cdot]$, with autocovariance functions satisfying
{\setlength{\belowdisplayskip}{5pt} \setlength{\belowdisplayshortskip}{5pt}
\setlength{\abovedisplayskip}{5pt} \setlength{\abovedisplayshortskip}{5pt}
\begin{alignat*}{2}
    C_{\tilde{s}}[n,l]&\triangleq\Expec{\tilde{s}[n]\,\tilde{s}[n+l]}, \;\; C_{\tilde{s}}[n,l]&=C_{\tilde{s}}[n+N_s,l],\\
    C_{\tilde{b}}[n,l]&\triangleq\Expec{\tilde{b}[n]\,\tilde{b}[n+l]}, \;\; C_{\tilde{b}}[n,l]&=C_{\tilde{b}}[n+N_b,l],
\end{alignat*}
}i.e., cyclostationary with fundamental periods $N_s, N_b > 1$ .

We denote the time offsets by $\tau_s$, $\tau_b$. Hence, 
{\setlength{\belowdisplayskip}{5pt} \setlength{\belowdisplayshortskip}{5pt}
\setlength{\abovedisplayskip}{5pt} \setlength{\abovedisplayshortskip}{5pt}
\begin{align*}
    \ust = [s_{\tau_s}[0], \hdots, s_{\tau_s}[N-1]]^{{\rm T}}\in \Cset^{N}, \; s_{\tau_s}[n] \triangleq \tilde{s}[n + \tau_s],
\end{align*}
}and similarly for $\ubt$ and $\myvec{\tilde{b}}$. We consider the case where time offsets are random, drawn from a discrete uniform distribution, i.e., $\tau_s\sim \mathcal{U}[0, N_s-1]$ and $\tau_b\sim \mathcal{U}[0, N_b-1]$, and assume $\tau_s$, $\tau_b$, $\tilde{s}[\cdot]$, $\tilde{b}[\cdot]$ and $\kappa$ are statistically independent. Consequently, note that $\ust$ and $\ubt$ are Gaussian mixtures.

As mentioned in Section \ref{sec:intro}, we are particularly interested in the case where we do not have full knowledge of the signal models---the distributions of $\tilde{\us}$ and $\tilde{\ub}$ (and thereby of $\ust$ and $\ubt$) are unknown. Nevertheless, we assume we have a dataset of $M$ independent, identically distributed (iid) copies of $\{(\uy^{(i)},\ust^{(i)})\}_{i=1}^M$, allowing for a data-driven approach. Note that this dataset contains \emph{unsynchronized} examples, wherein the corresponding time offsets $\tau^{(i)}_s$ and $\tau^{(i)}_b$ (of the $i$-th pair) are unknown.

Under this formulation, an ML approach is to learn a regressor, i.e., to estimate $\ust$ from $\uy$. Note that the dimension of the regression output is the same as that of the input. Thus, it can be thought of as a transformation or filtering (through some parametric model) of $\uy$ to obtain $\ust$, which is generally more challenging than a conventional single-output regression problem \cite{Baietto2022Radar}. We will thus investigate the use of deep neural networks (DNNs) to learn such a regressor from data. However, we begin with an in-depth examination of the SCSS problem through approaches to optimal estimation.

\vspace{-0.3cm}
\section{Optimal Model-Based Estimators}\label{sec:mmse}
\vspace{-0.3cm}
We now derive, for several cases, optimal estimators that achieve the lower bounds for their respective cases. Specifically, the simplified expression of the MMSE estimator reveals the challenges in attaining it, but also provides valuable intuition that informs and justifies our proposed data-driven methodology for the SCSS problem.

A possible ``classical signal processing" approach for this setting, is a two-step process, where one first estimates the model parameters, and then adopts MMSE estimation based on the empirical model. However, estimation of these parameters requires synchronization of the given dataset, which can be a very challenging task in itself. Nevertheless, in this section we assume access to an oracle regarding the signal model, so as to establish a lower bound on the mean squared error (MSE) for our problem.

Specifically, in the following we assume oracle knowledge of the signal models---i.e., the first and second-order statistics (SOSs) of $\myvec{\tilde{s}}$ and $\myvec{\tilde{b}}$, and the marginal distributions of $\tau_s,\tau_b$. 
We denote the conditional (temporal) covariance of $\ust$ given $\tau_s$ by
{\setlength{\belowdisplayskip}{5pt} \setlength{\belowdisplayshortskip}{5pt}
\setlength{\abovedisplayskip}{5pt} \setlength{\abovedisplayshortskip}{5pt}
\begin{align*}
    \mymat{C_s}(\tau_s) \triangleq \Expec{\ust \ust^{\her}|\tau_s}\in\Cset^{N\times N},
\end{align*}
}which we highlight that it is a function of $\tau_s$; likewise, $\mymat{C_b}(\tau_b)$ denotes the conditional covariance of $\ubt$. We denote the entries of $\mymat{C_s}(\tau_s)$ by $(\mymat{C_s}(\tau_s))_{i,j} = C_{\tilde{s}}[i+\tau_s, i-j]$. 

\vspace{-0.2cm}
\subsection{Case 1: The Optimal Linear MMSE Solution}
\vspace{-0.2cm}

To obtain the SOS of $\ust$, one must account for the uniform randomness in $\tau_s$. Hence, the covariance of $\ust$ is given by
{\setlength{\belowdisplayskip}{5pt} \setlength{\belowdisplayshortskip}{5pt}
\setlength{\abovedisplayskip}{5pt} \setlength{\abovedisplayshortskip}{5pt}
\begin{align*}
    \mymat{\check{C}_{s}}\hspace{-0.025cm}=\hspace{-0.025cm}\Expec{\ust \ust^{\her}}\hspace{-0.025cm}=\hspace{-0.025cm}\Expecto{\Expec{\ust \ust^{\her}|\tau_s}}{\tau_s}\hspace{-0.025cm}=\hspace{-0.025cm}\frac{1}{N_s}\sum_{\tau_s=0}^{N_s-1} \mymat{C_{s}}(\tau_s).
\end{align*}
}Note that this corresponds to a Toeplitz covariance structure, due to the fact that $s_{\tau_s}[\cdot]$ is a wide-sense stationary process (unlike $\tilde{s}[\cdot]$). The same applies for the covariance of $\ubt$.

The SOSs of $\ust$ and $\ubt$ are sufficient for optimal \emph{linear} estimation. For the sake of practicality, if we restrict our attention to linear operators, the optimal estimator is given by
{\setlength{\belowdisplayskip}{5pt} \setlength{\belowdisplayshortskip}{5pt}
\setlength{\abovedisplayskip}{5pt} \setlength{\abovedisplayshortskip}{5pt}
\begin{equation}\label{lmmse_est}
    \ustestsub{\LMMSE} = \mymat{\check{C}_{s}}  \, \left[\mymat{\check{C}_{s}}  + \Expecto{\kappa^2}{}\cdot \mymat{\check{C}_{b}}  + \sigma^2 \mymat{I}\right]^{-1} \uy,\\
\end{equation}
}where we used the statistical independence assumption of the sources and noise to compute the covariance of $\uy$. This would indeed correspond to the estimator that achieves the MMSE among the family of linear estimators. Further, given a sufficiently large dataset, one can implement an accurate approximation of \eqref{lmmse_est} by replacing the covariance matrices with empirical estimates (i.e., sample covariances) from the datasets. However, due to the random time offsets, $\ust$ and $\ubt$ are not Gaussian---but rather, are Gaussian mixtures---and hence the optimal linear estimator does not coincide with the MMSE estimator.

\vspace{-0.35cm}
\subsection{Case 2: The Oracle MMSE Solution}
\vspace{-0.2cm}

In this subsection, we develop the optimal solution for the case in which $(\tau_s,\tau_b,\kappa)$ are known (or, observable). This would have been the case if we would have at our disposal an oracle that provides side information of perfect synchronization to both the reference and interference signals, and the SIR-related coefficient $\kappa$. Under this setting, $\uy$ and $\ust$ are jointly Gaussian, hence the optimal solution is given by
{\setlength{\belowdisplayskip}{5pt} \setlength{\belowdisplayshortskip}{5pt}
\setlength{\abovedisplayskip}{5pt} \setlength{\abovedisplayshortskip}{5pt}
\begin{equation}\label{optimaloraclesolution}
    {\ustest}|_{(\tau_s, \tau_b, \kappa)} = \Expec{\ust|\uy, \tau_s, \tau_b, \kappa} = \mymat{H}(\tau_s, \tau_b, \kappa) \uy,
\end{equation}
}where
{\setlength{\belowdisplayskip}{5pt} \setlength{\belowdisplayshortskip}{5pt}
\setlength{\abovedisplayskip}{5pt} \setlength{\abovedisplayshortskip}{5pt}
\begin{align}\label{oraclefilter}
    \hspace{-0.05cm}\mymat{H}(\tau_s, \tau_b, \kappa)\hspace{-0.025cm}\triangleq\hspace{-0.025cm}\mymat{C_{s}}(\tau_s) \left[\mymat{C_{s}}(\tau_s) + \kappa^2\mymat{C_{b}}(\tau_b) + \sigma^2\mymat{I} \right]^{-1} 
\end{align}
}is the optimal linear (generally) time-varying filter. Since \eqref{oraclefilter} is a function of $(\tau_s, \tau_b, \kappa)$, we subsequently refer to \eqref{optimaloraclesolution} as an oracle-synchronized MMSE solution. Nevertheless, it should be noted that \eqref{optimaloraclesolution} is not a realizable estimator, as it is a function of latent, unobservable variables. Also, note that for cyclostationary signals, \eqref{optimaloraclesolution} can also be represented as a frequency-shift filter or a cyclic Wiener filter, i.e., expressed as a linear function of frequency shifted copies of the observation \cite{schlezinger2014fresh, gardner1994cyclostationarity}. 

\vspace{-0.2cm}
\subsection{Case 3: The Optimal MMSE Estimator}
\vspace{-0.2cm}

In general, the time offsets $\tau_s$ and $\tau_b$, and the SIR parameter $\kappa$, are not known at inference time, and a realizable estimator must account for their inherent randomness, by explicit/implicit (non-linear) estimation. However, upon conditioning on these quantities, i.e., considering \eqref{oraclefilter} as fixed, the resulting optimal MMSE estimator would also be the linear one. With this observation in mind, the true, realizable MMSE estimator is expressed as
{\setlength{\belowdisplayskip}{5pt} \setlength{\belowdisplayshortskip}{5pt}
\setlength{\abovedisplayskip}{5pt} \setlength{\abovedisplayshortskip}{5pt}
\begin{align}
    \ustestsub{\MMSE} &= \Expec{\ust|\uy} = \Expecto{\Expec{\ust | \uy, \tau_s, \tau_b, \kappa }}{(\tau_s, \tau_b, \kappa)|\uy} \nonumber \\
    &= \sum_{\tau_s=0}^{N_s-1}\sum_{\tau_b=0}^{N_b-1}\sum_{\kappa\in\mathcal{K}} p(\tau_s, \tau_b, \kappa|\uy) 
    \, \ustest|_{(\tau_s, \tau_b, \kappa)}, \label{mmse}
\end{align}
}which essentially corresponds to the sum of oracle-synchronized MMSE solutions~\eqref{optimaloraclesolution} for each set of parameters, weighted by the corresponding posterior probabilities given the observation $\uy$. We emphasize that \eqref{mmse}, unlike \eqref{optimaloraclesolution}, is a legitimate estimator, i.e., a function of the observed data only. However, the MSE achieved by \eqref{optimaloraclesolution} serves as a lower bound of that by \eqref{mmse}, by data procecessing inequality of MMSE.

It should be noted that, in cases when the true posterior is a Kronecker delta function, \eqref{mmse} in fact coincides with \eqref{optimaloraclesolution},
{\setlength{\belowdisplayskip}{5pt} \setlength{\belowdisplayshortskip}{5pt}
\setlength{\abovedisplayskip}{5pt} \setlength{\abovedisplayshortskip}{5pt}
\begin{align}
    p(\tau_s, \tau_b, \kappa|\uy) &= \delta[\tau_s-\tau^*_s, \tau_b-\tau^*_b, \kappa-\kappa^*]\label{deltaposterior} \\
    \Longrightarrow \;\; \ustestsub{\MMSE} &=  {\ustest}|_{(\tau^*_s, \tau^*_b, \kappa^*)},\nonumber
\end{align}
}namely the performance of the oracle-aided solution is achievable. Indeed, for the noiseless case ($\sigma^2=0$), it is easy to construct examples, where perfect separation is attained.

\begin{figure}
    \centering\vspace{-0.cm}
    \includegraphics[width=0.5\textwidth]{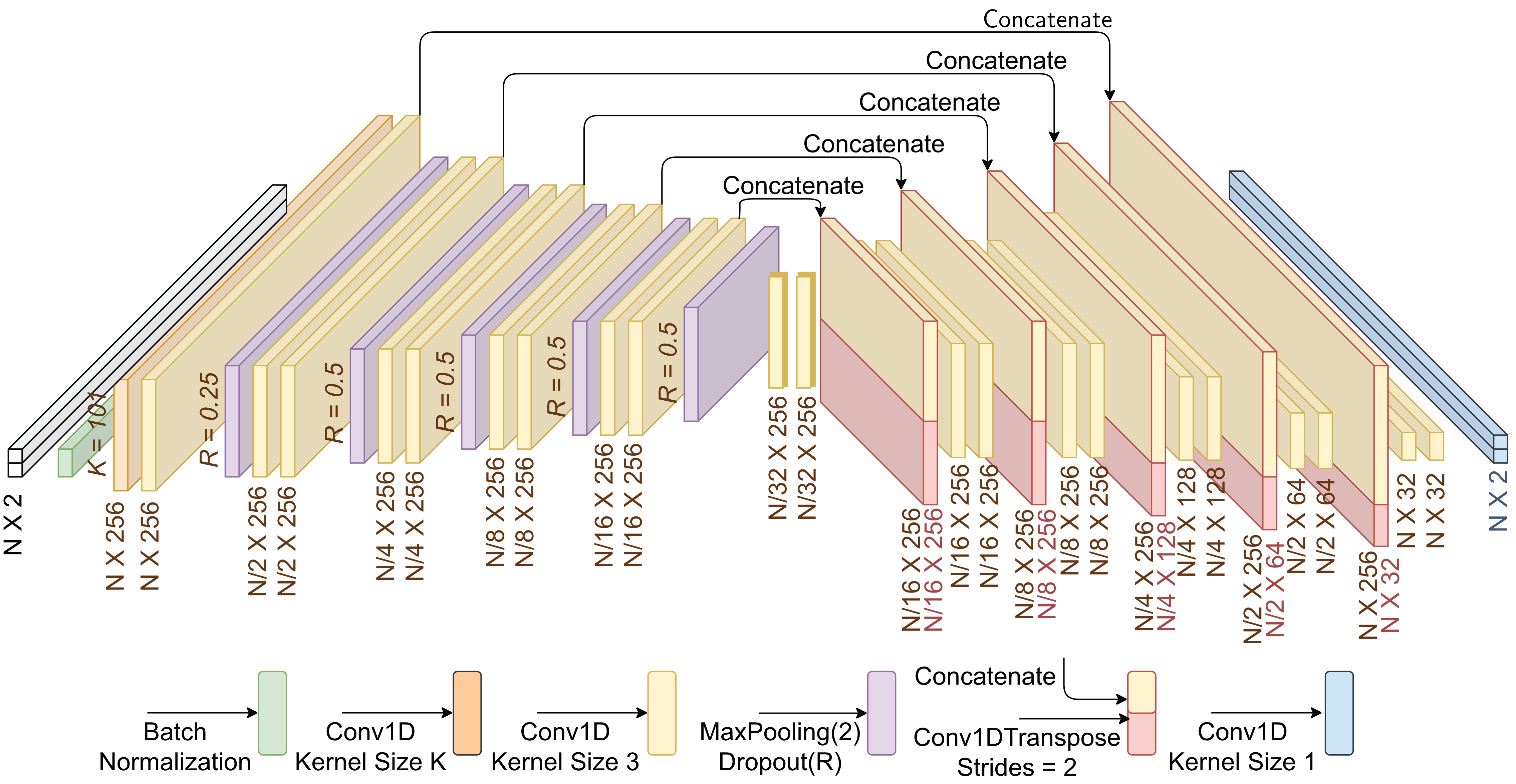}\vspace{-0.cm}
    \caption{The U-Net architecture used for SCSS in our simulations.}\vspace{-0.6cm}
    \label{fig:unet}
\end{figure}

\vspace{-0.3cm}
\section{Methodologies for Data-Driven SCSS}\label{sec:dnn}
\vspace{-0.2cm}
The methods discussed thus far provide an understanding of the functional structure for the MMSE estimator, in terms of the (unknown) time offsets and SIR parameters. Nevertheless, implementing the MMSE estimator may not be possible in practice, due to the challenges outlined below.

\vspace{-0.2cm}
\subsection{Difficulties in Realizing the MMSE Estimator}
\vspace{-0.2cm}
First, as seen in \eqref{mmse}, the MMSE estimator involves the computation of a posterior term over the latent variables $(\tau_s, \tau_b, \kappa)$. However, obtaining this posterior becomes computationally intensive as the space of parameters increases. 

Second, \eqref{mmse} also involves a sum of different (conditionally) linear operators, each corresponding to a set of parameter values. However, each of these operators involves an inversion of a large covariance matrix, which is infeasible in regimes of signals from long observation periods (large $N$). 

Lastly, and importantly, in practice we do not have oracle knowledge of the signal model, corresponding to the first and second-order statistics in the case of the Gaussian models. Instead, we are given many examples, through which we can obtain empirical estimates of the statistics. However, as mentioned, the dataset of the signals is not synchronized. Thus, obtaining the SOSs of the underlying cyclostationary signal, $\mymat{C_{\tilde{s}}}$ and $\mymat{C_{\tilde{b}}}$, requires the estimation of the latent variable $\tau_s$ and $\tau_b$, for each example in the dataset, corresponding to the synchronization of the entire dataset, which is generally an extremely challenging task of independent interest in itself.

We now show that ML methods can be successfully used to circumvent the aforementioned issues. Specifically, we propose the use of DNNs, trained on \emph{unsynchronized} datasets, to solve the SCSS problem. We use two representative examples to demonstrate our approach, and compare it with the performance of an optimal MMSE solution that utilizes oracle knowledge.  
The goal henceforth is to establish a practical pipeline, and benchmark it against a theoretical lower bound.

\vspace{-0.2cm}
\subsection{Supervised Separation with U-Net}
\vspace{-0.2cm}

Given the formulation in Section \ref{sec:prob}, a natural approach is to use a DNN to learn a regression model with multivariate output for source separation.
We propose to use the so-called ``U-Net'' architecture for SCSS (architecture shown in Fig.~\ref{fig:unet}).\footnote{\label{github} \url{https://github.com/RFChallenge/SCSS_CSGaussian}} Such a DNN was first proposed for biomedical image segmentation \cite{ronneberger2015unet}, but has found its use in other applications, including spectrogram-based RF interference cancellation \cite{akeret2017radio} and audio source separation \cite{stoller2018wave, tzinis2020sudo}---all of which also corresponding to a multivariate regression setup with the same dimensions on the input and output. Similar to the latter works, we use 1D-convolutional layers to capture features relating to the time series data. The U-Net architecture contains downsampling blocks that operate on successively coarser timescales, and possesses skip connections that combine features at these various timescales with the upsampling blocks. 

To technically handle complex-valued signals, borrowing inspiration from widely linear estimation \cite{picinbono1995widely}, we stack the real and imaginary parts as separate channels to the U-Net.

As these methods are applied to time series signals in practice, the use of domain knowledge to craft an appropriate neural network architecture may be crucial in attaining performance gain, as evident in our experiments and architectural choices. For example, we made the intentional choice of longer kernel sizes on the first convolutional layers. 
This further reinforces the relevance of this work, that is in identifying and characterizing DNN architectures under study relative to the best possible performance. In our experiments, we observe that kernel sizes that match the effective correlation length (i.e., timescales in which the covariance magnitudes are non-negligible) are required to attain the best performance. This may be an indication to how some partial, though important, information about the signal model is helpful (or even essential) in seeking the appropriate DNN architecture.

\vspace{-0.3cm}
\section{Simulation Results}\label{sec:expt}
\vspace{-0.2cm}

We now consider two examples for SCSS. For each setting, we train a U-Net to estimate the corresponding signal $\ust$ from the mixture $\uy$ in an end-to-end fashion, and with no supervision regarding the time-shifts $\tau_s$ and $\tau_b$, and the gain $\kappa$.

In the examples below, we describe how long (but finite) segments of the processes $\tilde{s}[\cdot], \tilde{b}[\cdot]$ are generated, from which $N$-length segments are extracted to create the datasets. The training set is processed as such to yield a labeled dataset of iid copies---mixture and ground-truth reference signal, $\{(\uy^{(i)},\ust^{(i)})\}_{i=1}^M$---as is done in the supervised learning framework. Our training set comprises $10,000\times |\mathcal{K}|$ pairs of mixtures $\uy$ and ground-truth $\ust$, and the validation set comprises $500\times |\mathcal{K}|$ pairs, where the cardinality $|\mathcal{K}|$ is the total number of levels for $\kappa$ under consideration. Subsequently, we 
test the performance across $1,000$ examples per $\kappa$ level, reporting the average MSE in dB. Note that varying $\kappa$ results in different levels of SIR. In our simulations, we assess the separation performance across different SIR levels. We also compare the performance of using the aforementioned U-Net against that of the optimal model-based estimators. For these optimal estimators, $\kappa$ is assumed to be known.

\vspace{-0.2cm}
\paragraph*{Implementation Details:}Keras and Tensorflow 2 are used to implement and train the U-Net \cite{chollet2015keras, tensorflow2015-whitepaper}. For training, we use empirical MSE as the loss function. We also use Adam optimizer\cite{kingma2017adam} and an exponentially decaying learning rate schedule, batch size of 32 with shuffled training samples, and trained for 2,000 epochs with early stopping if there is no improvement for 100 epochs on the validation set. 
We train the neural networks on a computing cluster with Intel Xeon Gold 6248, 192 GB RAM, and an NVidia Volta V100 GPU.

\vspace{-0.2cm}
\subsection{Signals with Randomly Generated Covariances}
\begin{figure}
    \centering
    \setlength\tabcolsep{2 pt}
    \begin{tabular}{c c||c c}
        \hline
        \multicolumn{4}{l}{(a) Simple Cyclostationary Example}\\\hline
        \includegraphics[width=0.11\textwidth]{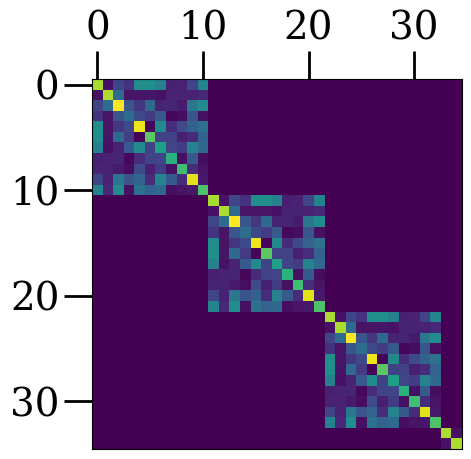} & 
        \includegraphics[width=0.11\textwidth]{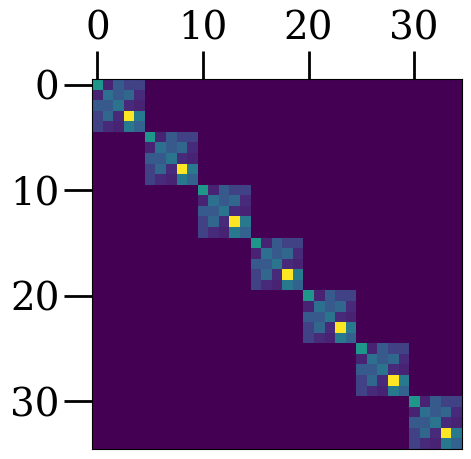} &
        \includegraphics[width=0.11\textwidth]{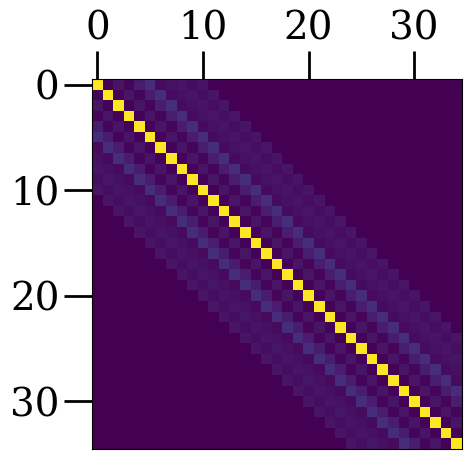} &
        \includegraphics[width=0.11\textwidth]{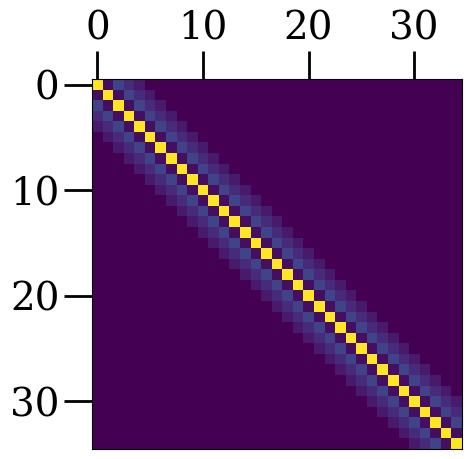} \\
        \footnotesize{Signal $\ust$} & \footnotesize{Signal $\ubt$} & \footnotesize{Signal $\ust$} & \footnotesize{Signal $\ubt$}\\[-0.1cm]
        \footnotesize{($\tau_s=0$)} & \footnotesize{($\tau_b=0$)} & \footnotesize{(Unsynchronized)} & \footnotesize{(Unsynchronized)} \\
        \hline
        \multicolumn{4}{l}{(b) Communication-inspired Example}\\\hline
        \includegraphics[width=0.11\textwidth]{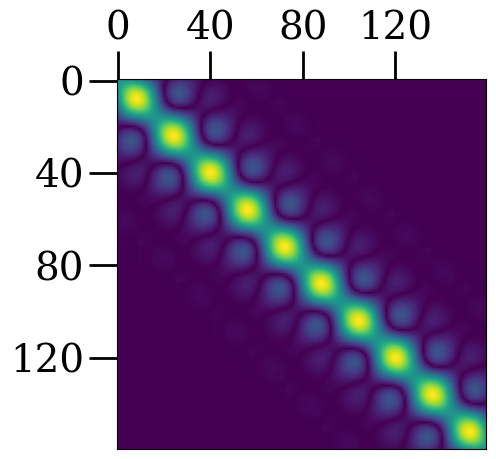} & 
        \includegraphics[width=0.11\textwidth]{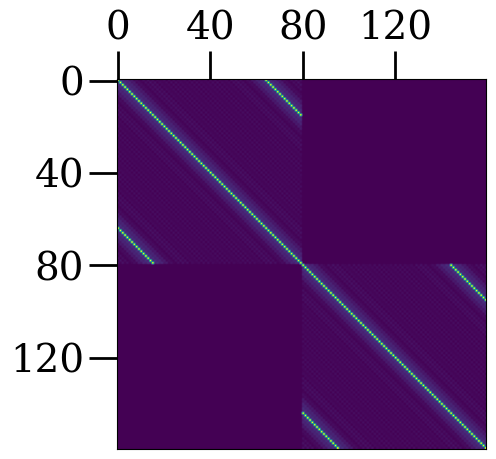} &
        \includegraphics[width=0.11\textwidth]{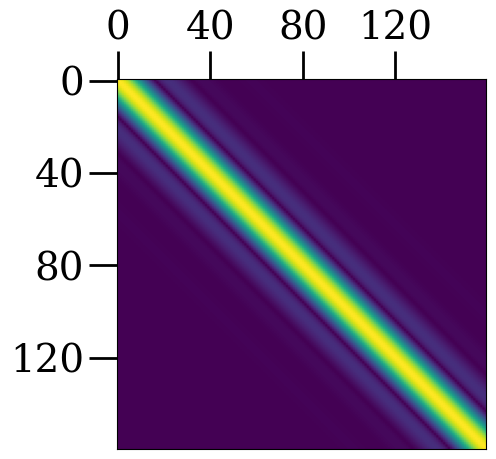} &
        \includegraphics[width=0.11\textwidth]{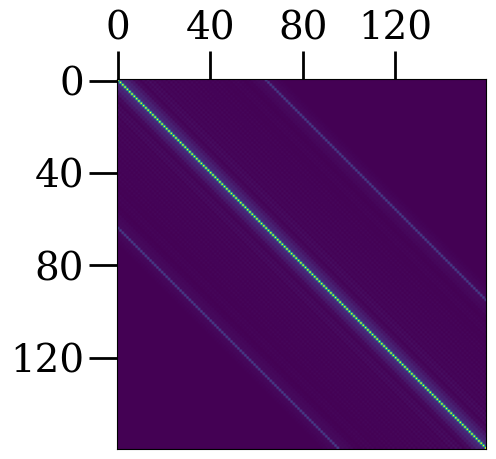} \\
        \footnotesize{RRC Signal} & \footnotesize{OFDM Signal} & \footnotesize{RRC Signal} & \footnotesize{OFDM Signal} \\[-0.cm]
        \footnotesize{($\tau_s=0$)} & \footnotesize{($\tau_b=0$)} & \footnotesize{(Unsynchronized)} & \footnotesize{(Unsynchronized)} \\
        \hline
    \end{tabular}\vspace{-0.1cm}
    \caption{Visualization of (a sub-matrix of) the covariance matrices for the signals used in the respective examples.}\vspace{-0.6cm}
    \label{fig:covariance}
\end{figure}

\vspace{-0.2cm}
We consider \eqref{mixture} with $N=256$, $N_s=11$, $N_b=5$, $\mathcal{K}$ corresponding to 5 equidistant SIR levels in $[-6,6]$ dB. The reference and interference signals are generated as
{\setlength{\belowdisplayskip}{5pt} \setlength{\belowdisplayshortskip}{5pt}
\setlength{\abovedisplayskip}{5pt} \setlength{\abovedisplayshortskip}{5pt}
\begin{equation*}
\myvec{\tilde{s}} = \mymat{G_s}\myvec{a}_1, \;  \; \myvec{\tilde{b}} = \mymat{G_b}\myvec{a}_2,
\end{equation*}
}where $\myvec{a}_1, \myvec{a}_2\sim \CNorm(\myvec{0}, \mymat{I})$, and $\mymat{G_s}, \mymat{G_b} \in \Cset^{\tilde{N} \times \tilde{N}}$, with $\tilde{N}=550$, are block-diagonal matrices with repeating $N_s \times N_s$ and $N_b \times N_b$ blocks respectively. Each entry in the blocks is drawn (once) independently from the Gaussian distribution, and is fixed for the rest of this experiment. Full details on the signal generation are provided in our Github repository.\footnotemark[2] Fig.~\ref{fig:covariance}(a) shows the covariance structures of the resulting sources.

\begin{figure}
    \centering
    \includegraphics[width=0.44\textwidth]{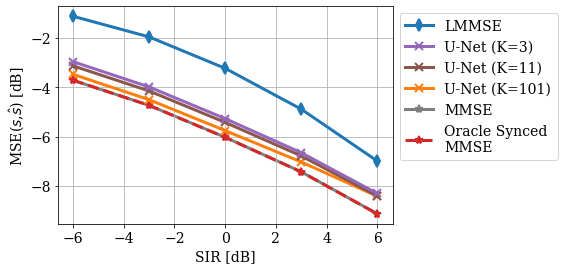} \vspace{-0.4cm}
    \caption{Separation performance of the U-Net separator vs.\ optimal model-based estimators for waveforms with randomly generated covariance structures.} \vspace{-0.3cm}
    \label{fig:results_cyclo}
\end{figure}

Fig.~\ref{fig:results_cyclo} compares the MSE achieved by the U-Net against that obtained by the linear and the ``global" MMSE estimators. We also include the oracle-synchronized MMSE~\eqref{optimaloraclesolution}, which, as evident from the figure, is indistinguishable (in terms of its MSE) from the true, nonoracle MMSE estimator \eqref{mmse}. This occurs when all the mass of the posterior is (approximately) concentrated at the point of the true values of $(\tau_s, \tau_b, \kappa)$, as in \eqref{deltaposterior}.

We observe that a U-Net trained on an unsynchronized dataset of signals is capable of obtaining results close to the MMSE performance. This means that the U-Net necessarily learned a significant part of the model, which enables high-quality estimation of the reference signal. We reiterate that the U-Net did not have access to the true statistics of the signal model or any form of synchronization of the signals during training and inference. The slight deterioration performance could be attributed to approximation errors introduced from a DL-based function approximator, or due to the trade-off from lack of access to synchronized dataset. Future work entails identifying factors to close the performance gap. 

\vspace{-0.2cm}
\subsection{Communication-like Waveforms}
\vspace{-0.2cm}

The problem of SCSS is particularly relevant in the application of wireless communications, where we may be interested in separating multiplex of signals, or extracting a signal-of-interest while mitigating interference \cite{rfchallenge, zhao2021scbss}. In this example, we consider two types of signals with SOS properties identical to a single-carrier communication waveform and an orthogonal frequency division multiplexed (OFDM) waveform, respectively.
The single-carrier signal is modeled as,
{\setlength{\belowdisplayskip}{5pt} \setlength{\belowdisplayshortskip}{5pt}
\setlength{\abovedisplayskip}{5pt} \setlength{\abovedisplayshortskip}{5pt}
\begin{align}
    &\tilde{s}[n] = \sum_{p=-\infty}^{\infty} a_p \, g[n - p N_s], \label{rrc}
\end{align}
}where $a_p \sim \CNorm(0,1)$, $N_s$ is the symbol period, and $g[n]$ is the root-raised cosine (RRC) filter. Here, we consider a RRC filter that corresponds to 16 samples per symbol---i.e., $N_s=16$---and spans 8 symbols, with a roll-off factor of 0.5. This also corresponds to an example where a cyclostationary signal's covariance (Fig.~\ref{fig:covariance}(b)) is not block-diagonal, which leads to additional computational burden in the matrix inversion.

\begin{figure}
    \centering
    \includegraphics[width=0.44\textwidth]{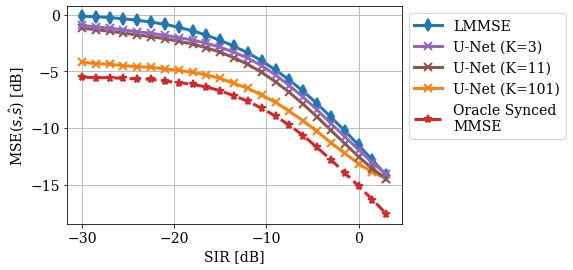} \vspace{-0.4cm}
    \caption{Separation performance of the U-Net separator vs.\ optimal model-based estimators for communication-like waveforms.} \vspace{-0.6cm}
    \label{fig:results_comm}
\end{figure}

The second source, an OFDM waveform, is modeled as,
{\setlength{\belowdisplayskip}{3pt} \setlength{\belowdisplayshortskip}{3pt}
\setlength{\abovedisplayskip}{2pt} \setlength{\abovedisplayshortskip}{2pt}
\begin{align}
    &\tilde{b}[n] = {\frac{1}{\sqrt{N_{\textrm{sc}}}}}  \sum_{p=-\infty}^{\infty} \sum_{\ell=0}^{N_{\textrm{sc}}-1} a_{p,\ell} \, q[n - p N_b, \, \ell], \label{ofdm} \\[-0.2cm]
    &q[n, \ell] = \mathbbm{1}_{\left\{0 \leq n \leq N_b-1\right\}} \, \cdot \,  \exp\left(j 2\pi \ell \frac{n - N_{\textrm{cp}}}{N_{\textrm{sc}}}\right), \nonumber
\end{align}
}where $a_{p,\ell} \sim \CNorm(0,1)$ for $\ell\in\mathcal{L}_{\textrm{sc}}$, where $\mathcal{L}_{\textrm{sc}}$ refers to the set of nonzero subcarrier indices, and $a_{p,\ell}=0$ otherwise, $N_{\textrm{sc}}$ is the number of subcarriers per OFDM symbol, $N_{\textrm{cp}}$ is the cyclic prefix (CP) length, and $N_b$ is the OFDM symbol period, i.e., $N_b = N_{\textrm{sc}} + N_{\textrm{cp}}$. 
In our specific example,
$N_{\textrm{sc}}=64$, $N_{\textrm{cp}}=16$, and thus $N_b=80$. More details on the signal specifications are provided in our Github repository.\footnotemark[2] Fig.~\ref{fig:covariance}(b) shows the covariance structures of the resulting sources, whose cyclostationarity is evident.

We consider segments of length $N=1280$, and $\mathcal{K}$ corresponding to SIR levels $[-30,3]$ dB at 1.5 dB steps ($|\mathcal{K}|=23$), namely we focus on the more challenging low SIR regime. 
We emphasize that the models \eqref{rrc} and \eqref{ofdm} are assumed to be \emph{unknown} once we have generated the dataset. Rather, we only have access to a dataset of unsynchronized samples.

Fig.~\ref{fig:results_comm} compares the MSE achieved by the linear MMSE and the oracle-synchronized MMSE.
We note that, for this example, the true MMSE curve could not be obtained in practice due to the size of the parameter space, rendering the computation of the posterior infeasible. Nevertheless, the MSE of the oracle-synchronized MMSE solution---albeit not a realizable estimator, as established earlier---serves as a lower bound.

As observed from Fig.~\ref{fig:results_comm}, the best performing U-Net, which is trained on unsynchronized data, outperforms the linear MMSE estimator, and is close to the performance of the oracle MMSE (e.g., about $1.2$ dB away at SIR levels between $-9$ and $-30$ dB). We also highlight that the choice of U-Net architecture to achieve such a performance benefits from specific domain knowledge. For example, capturing the temporal structures on the order of the signals' effective correlation length yielded significantly improved performance---for which long kernels on the first layer is a way of doing so.\footnote{For comparisons with other DNNs for the communication example: \hspace{1cm} \url{https://github.com/RFChallenge/SCSS_DNN_Comparison}}

% In practice, digital communication waveforms tend to depart from Gaussianity; in particular, they possess discrete constellation symbol sets, inducing a stronger structure on the latent sources. Hence, the MMSE estimator in these cases is no longer linear that solely depends on first and second-order statistics. In fact, performance better than what is attained by \eqref{mmse} can be achieved by capturing and exploiting higher-order statistics. Nevertheless, an analytical form of such optimal MMSE estimators is difficult to obtain. This is exactly the point where learning based method can come into play, as they hold the promise of improving upon optimal linear estimators by learning structures relating to such higher-order statistics. Characterization and exploration of such gains are of high interest for future work. 

\vspace{-0.3cm}
\section{Conclusion}\label{sec:conclusion}
\vspace{-0.25cm}

We study the SCSS problem when signal models of the underlying latent sources are unknown, and develop a data-driven, DL approach as a prospective solution. Further, we characterize the setup for signals corresponding to segments extracted from cyclostationary Gaussian processes, and present optimal MMSE estimators. We then outline practical challenges associated with implementing the MMSE estimator, and through two examples, demonstrate how a trained U-Net with a careful choice of first-layer kernel size is able to achieve comparable performance, motivating DNNs as a competitive, computationally attractive approach. 

We reiterate that in this work, the Gaussianity assumption allowed us to have a simplified closed-form expression for the MMSE estimator~\eqref{mmse}, and in turn a theoretical benchmark to compare against. 
In practice, signals depart from Gaussianity---e.g., digital communication signals possess symbols from discrete sets. In these cases, the MMSE estimator is no longer linear that solely depends on (up to) second-order statistics, and higher-order statistics can be exploited to attain performance better than what is attained by \eqref{mmse}. However, an analytical form of the optimal estimator is less obvious. This strongly motivates learning-based methods that hold the promise of improving upon implementable, optimal model-based estimators by learning those higher-order structures. Characterization and exploration of such performance gains via DL are of high interest for future work. 

\vspace{-0.3cm}
\bibliographystyle{IEEEtran}
\small{\bibliography{refs}}

\end{document}